\documentstyle{EuroPhys}
\input EuroMacr
\input epsf
\input isolatin1.sty
\begin{document}
\newcommand{\io}
   {\mathrel{\rlap{\raise1pt\hbox{$<$}}{\lower4pt\hbox{$\sim$}}}}     
\euro{}{}{}{}
\Date{}
\shorttitle{}

\title{Long-range Casimir interactions between impurities in nematic
liquid crystals and the collapse of polymer chains in such solvents}
\author{D. Bartolo, D. Long \cite{newaddress}, and J.-B. Fournier}
\institute{Laboratoire de Physico-Chimie Th{\'e}orique\\
  {\'E}cole Sup{\'e}rieure de Physique et de Chimie Industrielles de la Ville
  de Paris\\ 10 rue Vauquelin, F-75231 Paris C{\'e}dex 05, France}
\rec{}{}

\pacs{ \Pacs{}{}{} \Pacs{}{}{} } 
\maketitle

\begin{abstract}
The elastic interactions between objects embedded in a nematic liquid
crystal are usually caused by the average distorsion---rather than by
the fluctuations---of the nematic orientational field. We argue that for
sufficiently small particles, the nematic-mediated interaction
originates purely from the fluctuations of the nematic director. This
Casimir interaction decays as $d^{-6}$, $d$ being the distance between
the particles, and it dominates van der Waals interactions close to the
isotropic--to--nematic transition.  Considering the nematic as a polymer
solvent, we show that the onset of this Casimir interaction at the
isotropic--to--nematic transition can discontinuously induce the
collapse of a flexible polymer chain from the swollen state to the
globular state, without crossing the $\Theta$-point.
\end{abstract}
In $1948$, Casimir originally showed that the interplay between
electromagnetic quantum fluctuations and boundary conditions generates
long-range interactions between uncharged conducting
walls~\cite{casimir}. The corresponding energy density per unit area
falls off as $d^{-3}$, where $d$ is the distance between the two walls.
Similar effects can also arise in liquid crystals~\cite{ajdari,ajdari2}.
Nematic liquid crystals are fluid mesophases composed of rodlike
molecules exhibiting a broken rotational symmetry along a nonpolar
direction, which is characterized by a unit vector ${\bf n}$ called the
nematic director~\cite{degennes}.  Although a nematic material placed
between two walls imposing identical orientational boundary conditions
does not store any elastic energy, and hence does not give rise to any
mean-field force between the walls, it was shown that the interplay
between the director's thermal fluctuations and the boundary conditions
induces a ``Casimir'' long-range interaction falling off as
$d^{-2}$~\cite{ajdari2}. Note that Casimir-like interactions 
have been reviewed recently in \cite{kardar}. 

Such Casimir interactions should exist without any distortion of the
average director orientation, which makes them so intriguing. However,
they have not yet been experimentally brought to the fore. Indeed, in
most studied systems, e.g.\ macroscopic colloidal particles immersed in
nematic liquid crystals~\cite{poulin1,poulin2}, the interactions
mediated by the nematic phase are dominated by the energy of the
mean-field elastic distorsions of the director field, which decays as
$d^{-3}$ or $d^{-5}$ depending on the symmetries of the
inclusions~\cite{poulin1,terentjev,lev}. We argue here that inclusions
of size smaller than the molecules's orientational correlation
length~$\xi$ should experience pure Casimir interactions.  The latter
are found to decay as $d^{-6}$, and should dominate van der Waals
interactions in the vicinity of the nematic--to--isotropic transition. 
Then, considering a single flexible polymer chain in a nematic
solvent~\cite{cotton}, we show that this Casimir interaction can induce
the collapse of the chain at the isotropic--to--nematic transition.

In a nematic liquid crystal, the constituent molecules are locally
oriented on the average parallel to some direction~${\bf n}$. Close to
the isotropic transition, although the macroscopic degree of order can
be very weak, the molecular orientations are correlated over
``swarms'' of size~$\xi$, which can be as large as $10$ to $20$ times
the molecular dimensions~\cite{degennes}. On length-scales larger than
$\xi$, these essentially uncorrelated swarms fluctuate about their
common orientation~${\bf n}$, the fluctuations being all the more
important than the degree of order is small.  To construct the nematic
order-parameter ${\bf Q}$---a traceless tensor describing the amount of
uniaxial anisotropy of the nematic phase~\cite{degennes}---one can
average over a small volume the microscopic quantity ${\bf
q}=\frac{3}{2}({\bf m}{\bf m}-\frac{1}{3}{\bf I})$ made from the
molecular orientation ${\bf m}$ and the identity tensor~${\bf I}$.
However, it follows from the previous discussion that the averaging
volume must be of dimension larger than $\xi$ for this average to be
representive of the macroscopic degree of order. One actually defines
the director ${\bf n}$ and the amount of order $S$ by writting the
average of ${\bf q}$ in the form ${\bf Q}=\frac{3}{2}S({\bf n}{\bf
n}-\frac{1}{3}{\bf I})$. Hence, both~$S$ and~${\bf n}$ are only defined
on length-scales larger than the cutoff~$\xi$.

\begin{figure} \centerline{\epsfxsize=5cm\epsfbox{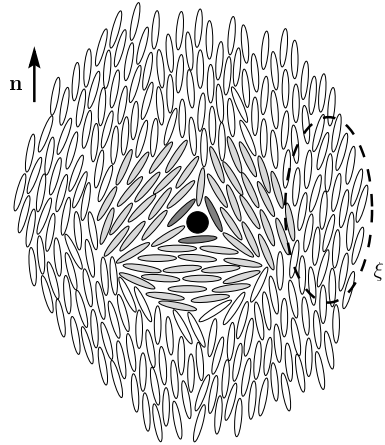}}
\caption{Snapshot of a nematic liquid crystal in the vicinity of a
disordering impurity (black sphere).  The nematic molecules under the
direct influence of the impurity are drawn in dark gray.  The
correlation ``swarms'', of size $\xi$, indirectly influenced by the
impurity are drawn in light gray. Note that the orientational
fluctuations within the swarm have been diminished for the sake of
clarity. The director ${\bf n}$, which is only defined by spatial
averages over distances larger than $\xi$ is not influenced by the
impurity. In particular, the above $+1/2$ and $-1/2$ disclinations of
the molecular orientation are irrelevant, since their separation
compares with $\xi$.  However, after coarse-graining on a scale larger
than $\xi$, the impurity is seen to reduce the nematic degree of order
$S$, thereby inducing a point-like reduction of the Frank elastic
constants.}

\label{dessin}
\end{figure}

Let us now consider a microscopic (neutral) impurity of size less than
$\xi$, embedded in the nematic phase. On the range of molecular forces
(a few molecular lengths), it will {\em directly\/} affect the
orientation of the nematic molecules, either reducing or enhancing their
parallelism~(see Fig.~\ref{dessin}). Due to the orientational
correlations of the nematic molecules, this direct influence will spread
over~$\xi$. However, the nematic director---which is only defined at
length-scales larger than~$\xi$---will not be affected by the presence
of the particle. This is because the orientational correlations
essentially vanish beyond~$\xi$~\cite{Goldstone}. Such a microscopic
impurity cannot therefore induce any mean-field distorsion of~$\bf n$,
as a macroscopic colloidal particle would (the difference being that the
latter is physically correlated on a macroscopic scale). Nevertheless, a
microscopic impurity should affect the local {\em degree of order\/}
$S$, because the swarms around it are strongly disoriented.  Hence,
within the macroscopic Landau-de Gennes description of nematics---which
is naturally coarse-grained over $\xi$---molecular impurities are simply
represented by point-like variations of the degree of order $S$.
Moreover, since the Frank elastic constants~\cite{Frank} associated with
the distorsions of the director field are proportional to
$S^2$~\cite{degennes}, such impurities should be modelled as point-like
variations of the Frank elastic constants.

To simplify, we describe the nematic elasticity in the one constant
approximation~\cite{degennes}, and we assume that the director
undergoes only small deviations with respect to ${\bf e}_z$.  Setting
${\bf n}\simeq(n_x,n_y,1)$ and considering a spatially varying elastic
constant
\begin{equation}
K({\bf r})=K+\delta\!k_1\,\delta({\bf r}-{\bf r}_1)
+\delta\!k_2\,\delta({\bf r}-{\bf r}_2),
\end{equation}
which models two molecular impurities located at ${\bf r}_1$ and
${\bf r}_2$, the system's free energy takes the form
\begin{equation}\label{H}
{\cal H}=\int\!d{\bf r}\,
\frac{1}{2}K({\bf r})\left[
\left(\nabla n_x\right)^2+\left(\nabla n_y\right)^2\right].
\end{equation}
Throughout the paper, all the energies are expressed in units of $k_{\rm
B}T$.  The minimum of~(\ref{H}) corresponds to the uniform state
$n_x=n_y=0$.  Therefore the inclusions do not experience any mean-field
interaction.  The total free energy of the system, however, which takes
into account the fluctuations of the director field, depends in a
non-trivial way on ${\bf r}=|{\bf r}_2-{\bf r}_1|$. Because $n_x$ and
$n_y$ are decoupled and play a symmetric role, the partition function
$Z$, and hence the total free energy, can be expressed as a functional
integration over $n_x$ only.  By performing a Hubbard-Stratonovich
transformation \cite{lub}, we obtain
\begin{eqnarray}
\hspace*{-1cm}&&Z^\frac{1}{2}\!=\!\int{\cal D}[n_x]\,\exp\Big\{\!-\!\!
\int\!d{\bf r}\,\big[
\frac{1}{2}K(\nabla n_x)^2
+\frac{1}{2}\sum_{j=1,2}\delta\!k_j\,
\delta({\bf r}-{\bf r}_j)\,(\nabla n_x)^2
\big]\Big\}\nonumber\\
\hspace*{-1cm}&&=\!\!
\int\!d{\bf h}_1d{\bf h}_2\,
e^{\displaystyle-\frac{h_1^2}{2\,\delta\!k_1}\!
-\!\frac{h_2^2}{2\,\delta\!k_2}}
\!\!\int\!{\cal D}[n_x]\,\exp\Big\{\!-\!\!
\int\!d{\bf r}\,\big[
\frac{1}{2}K(\nabla n_x)^2
+i\!\sum_{j=1,2}{\bf h}_j\!\cdot\!\nabla n_x\,\delta({\bf r}-{\bf r}_j)
\big]\Big\}.
\end{eqnarray}
Then, calculating all the Gaussians integrals yields
\begin{equation}
Z({\bf r})\propto{\rm det}^{-1}\pmatrix{
\displaystyle\frac{\bf I}{\delta\!k_1}-\nabla\nabla G({\bf 0})
&-\nabla\nabla G({\bf r})
\cr -\nabla\nabla G({\bf r})
& \displaystyle\frac{\bf I}{\delta\!k_2}-\nabla\nabla G({\bf 0})},
\end{equation}
in which the above $6\times6$ matrix, made of four $3\times3$ blocks,
involves the identity tensor ${\bf I}$, and the tensor $\nabla\nabla
G({\bf r})$. Here, $G({\bf r})$ is the Green function of the operator
$-K\nabla^2$.  The total free energy of the system, consisting of the
nematic liquid and of the two impurities, can then be calculated as
$F({\bf r})=-\ln Z({\bf r})$.  This total free energy can be decomposed
as the sum of the free energies of individual impurities in the nematic
solvent (which we call the ground-state free energy), and of the
interaction free energy between the two impurities.  In the following,
we consider only the latter, which we denote by $F_{\rm C}({\bf r})$ and
which goes to zero when the separation of the particles goes to
infinity.  The ground-state free energy appears as a constant that we
drop. 

To calculate the interaction free energy $F_{\rm C}({\bf r})$, one has
to introduce a short-distance cutoff $\Lambda^{-1}$.  Note that this
cutoff is arbitrary~\cite{ren}, however it cannot be smaller than the
correlation length $\xi$ of the nematic solvant, since the
free-energy~(\ref{H}) is not defined below shorter distances. After
regularization involving the cutoff $\Lambda$ in the reciprocal space,
we obtain (in units of
$k_{\rm B}T$)
\begin{equation}
F_{\rm C}(r)=
-\frac{27}{8\,\pi^2}\,
\frac{\delta\!K_1(\Lambda)\,\delta\!K_2(\Lambda)}
{\big[3K+\delta\!K_1(\Lambda)\big]
{\big[3K+\delta\!K_2(\Lambda)\big]}}
\left(\frac{\Lambda^{-1}}{r}\right)^6
+\,{\cal O}\!\left(r^{-12}\right)\,,
\end{equation}
with $\delta\!K_{i}(\Lambda)=(2\pi)^3\Lambda^{-3}\delta k_i(\Lambda)$.
The perturbations $\delta k_i(\Lambda)$ depend on the cutoff; this
dependence may be determined from the property that the interaction
free-energy, however, should not depend on the cutoff. Anyhow, when
$\Lambda^{-1}$ is chosen to be of order $\xi$, the perturbation of the
nematic order due to the impurity is expected to be of the same order as
the nematic order itself. Thus, from now on, we set $\Lambda=2\pi/\xi$
and we assume $|\delta\!K_{i}|\io K$.  

We find that the Casimir interaction is attractive when the two
inclusions are of the same nature, i.e.\ when
$\delta\!K_1\,\delta\!K_2>0$, and repulsive otherwise. Let us compare it
with the van der Waals interaction, which decays with the same power
law. With $F_{\rm vdW}(r)=-A/\pi\,\rho^2r^6$, in which $A$ is the
Hamaker constant and $\rho$ the inverse of the molecular volume of the
impurities~\cite{Isr}, we obtain
\begin{equation}
\frac{F_{\rm C}}{F_{\rm vdW}}=
\frac{27}{8\pi A}\left(\frac{\delta\!K}{3K+\delta\!K}\right)^2
\left(\rho\,\xi^3\right)^2\equiv\mu\left(\rho\,\xi^3\right)^2.
\label{casi}
\end{equation}
With typically $A\simeq1$ (in units of $k_{\rm B}T$) and $|\delta\!K|\io
K$, we expect $\mu$ in the range $10^{-2}$--$0.25$ (the latter value
corresponding to $\delta\!K\simeq-K$). Deep in the nematic phase, where
$\rho\,\xi^3\simeq10$, the Casimir and van der Waals interactions should
therefore be comparable. However, in the vicinity of the
nematic--to--isotropic transition, where $\rho\,\xi^3\simeq10^3$, the
Casimir interaction should be the leading one. There can be dramatic
consequences to this fact, as we argue now by considering a flexible
polymer embedded in a liquid crystal solvent~\cite{degennes2,Brochard}.

For flexible polymers, the persistence length corresponds to one monomer
length and is about 5\AA, which is smaller than the correlation
length~$\xi$ of the nematic liquid crystal. Thus, as discussed above,
the polymer cannot induce director gradients on the scale of one
persistence length (as a particle larger than~$\xi$ would, e.g.\ by
means of anchoring interactions). Beyond the chain persistence length,
the monomers orientations are uncorrelated, or, if they are correlated
\cite{trian}, this correlation is induced by the surrounding nematic
field, and cannot therefore induce any director gradient. We conclude
that the presence of a non-collapsed flexible polymer in a nematic
liquid does not give rise to Frank elasticity. In the following, we
shall neglect the possible large scale anisotropy of the Gaussian
conformation of the polymer induced by the nematic solvent.    

We suppose that the liquid crystal in its isotropic phase is a good
solvent for the considered flexible polymer. This property is
characterised by a positive excluded volume parameter $v_0$, which takes
into account the short-range microscopic monomer--monomer,
nematogen--monomer, and nematogen-nematogen interactions. The Flory
theory predicts a collapse of the polymer chains when the excluded
volume is negative. In the particular case $v_0=0$, called the
$\Theta$-point~\cite{degennes3}, the conformation of the chain is
essentially Gaussian. For positive values of the excluded volume, chains
are swollen as compared to the Gaussian random walk, and for chains long
enough, the end-to-end distance of the polymer is given, in the Flory
picture, by~\cite{degennes3}
\begin{equation}
\langle R^2\rangle^{\frac{1}{2}} \sim b^{\frac{2}{5}}
v_0^{\frac{1}{5}} N^{\frac{3}{5}},
\end{equation}
where $b$ is the monomer length and $N$ is the degree of polymerization
of the chain.  

The issue we address now is the effect of the isotropic--to--nematic
transition on the chain conformation, in the case of a flexible chain
with persistence length $b<\xi$. In the nematic phase, the effective
interaction between monomers results now from a bare excluded volume
$v_b$ that takes into account the short-range interactions, plus the
long-range Casimir interaction calculated previously. Since at
length-scales smaller than $\xi$, the nematic and the isotropic phases
are essentially indistinguishable from the point of view of the local
molecular organisation (as argued above), we do not expect the
isotropic--to--nematic transition to shift the value of the bare
excluded volume: thus we take $v_b \equiv v_0$.  The Casimir
interaction, however, is defined only on length-scales larger than
$\xi$. Hence, to compare the bare excluded volume and the Casimir
interaction, one must coarse grain the polymer in order that the monomer
length becomes $\xi$~\cite{charge}. This coarse-graining of the chain
imposes to rescale the bare excluded volume $v_0$. Thus, we redefine the
monomer according to
\begin{eqnarray}
&&N\to N/\lambda \equiv N', \nonumber \\
&&b\to b\,\lambda^{\nu} \equiv b',
\end{eqnarray}
while the end-to-end distance, which is unchanged, is now given by
\begin{equation}
\langle R^2\rangle ^{\frac{1}{2}} \sim b'^{\frac{2}{5}} v_0'^{\frac{1}{5}}
N'^{\frac{3}{5}},
\end{equation}
where $v'_0$ is the rescaled excluded volume. Thus, one has
\begin{equation}
v_0' = v_0\,\lambda^\gamma,
\end{equation}
with $\gamma =9/5$ if we take for $\nu$ the Flory exponent $\nu = 3/5$,
or $\gamma = 2$ if we assume that the chain has a Gaussian structure at
small scale, which corresponds to $\nu = 1/2$.  Note that rederiving the
end-to-end distance by the Flory argument---i.e.\ calculating a Gaussian
entropy for the chain, neglecting correlations to calculate the excluded
volume interaction, and balancing both contributions---yields the result
$\gamma=2$. Considering now monomers of size $b'=\xi$, we take $\lambda
= (\xi/b)^{\frac{1}{\nu}}$, and thus the rescaled excluded volume is
\begin{equation}
v_0' = v_0 \left(\frac{\xi}{b}\right)^{\frac{\gamma}{\nu}},
\end{equation}
with $\gamma/\nu=3$ when assuming a swollen state at small scale, or
$\gamma/\nu=4$ when assuming a Gaussian conformation at small scale. If
one considers chains not far from the $\Theta$-point, since $\xi$ is not
order of magnitudes larger than $b$, we can assume a Gaussian behaviour
at scales smaller than $\xi$~\cite{degennes3}.  A rescaling exponent $4$
is thus more reasonable than $3$, and will be retained for the
discussion below. 

The chain can now be seen as an ensemble of monomers of size~$\xi$,
submitted to the bare rescaled exluded volume interaction $v_0'$, and to
a long-range Casimir interaction characterized in our model by
$\delta\!K \simeq K$.  Note that the Casimir effect is pairwise additive
at lowest order. The effective excluded volume interaction is $v_{\rm
eff}=v'_0+v_{\rm C}$, with
\begin{equation}
v_{\rm C}=
\int_{r>\xi}\!\!d{\bf r}\,F_{\rm C}(r)=
-\frac{9}{2\pi}\left(\frac{\delta\!K}{3K+\delta\!K}\right)^2\!\xi^3
\equiv-\alpha\,\xi^3.
\end{equation}
This Casimir contribution is indeed attractive, and its effect is to
reduce the quality of the solvent. Adding the two contributions, we
obtain
\begin{eqnarray}
v_{\rm eff} = v_0\left(\frac{\xi}{b}\right)^4 - \alpha\,\xi^3,
\end{eqnarray}
with $\alpha\sim0.1$ for $\delta\!K\simeq K$. The collapse condition
$v_{\rm eff}<0$ can be written as
\begin{equation}
\frac{v_0}{b^3}\io\alpha\frac{b}{\xi}.
\end{equation}
Assuming $b/\xi\sim0.1$, we find that an abrupt collapse may be induced
by the isotropic--to--nematic transition, provided the isotropic phase
is not too far from the $\Theta$-point. For instance, with
$v_0/b^3\simeq10^{-2}$, chains with degree of polymerization $N>10^4$,
which are thus swollen in the isotropic phase~\cite{degennes3}, should
effectively collapse at the transition.  Note that the transition from
the swollen to the collapse state occurs without crossing the
$\Theta$-point: this is a consequence of the first-order nature of the
isotropic--to--nematic transition, which induces a discontinuous change
of the interactions between the monomers. One could not expect this
effect with a continuous second-order phase transition.

In conclusion, we have shown that small impurities in nematic solvents
undergo long-range Casimir interactions decaying as $d^{-6}$. For such
inclusions, elastic interactions are absent, while Casimir
interactions dominate over the van der Waals interactions. These
Casimir forces are attractive for identical particles and can lead to
the collapse of flexible polymer chains. They could thus be measured
by determining the phase diagram of various flexible polymers close
to the isotropic--nematic transition.

\stars
We thank A. Ajdari and L. Peliti for useful discussions.

%%%%%%%%%%%%%%%%%%%%%%%%%%%%%%%%%%%%%%%%%%%%%%%%%%%%%%%%%%%%%%%%%%%%%%


\begin{thebibliography}{99}
\bibitem{ren}The Frank elastic constants depend however weakly on this
cutoff.
\bibitem{charge}Constituents of the chain separated by a distance
smaller than $\xi$ can be considered as contributing to the ``charge''
$\delta K$ associated to the corresponding swarm of size $\xi$.
\end{thebibliography}
\end{document}